\documentclass{iaus}
\usepackage{graphicx}

\title[Optical pulsations from isolated neutron stars] 
{Optical pulsations from isolated neutron stars}

\author[Roberto P. Mignani]   
{Roberto P. Mignani$^{1,2}$
\affiliation{$^1$ Mullard Space Science Laboratory, University College London, \\ Holmbury St. Mary, Dorking, Surrey, RH5 6NT, UK \\email: {\tt rm2@mssl.ucl.ac.uk} \\ [\affilskip]$^2$ Kepler Institute of Astronomy, \\ University of Zielona G\'ora, Lubuska 2, 65-265, Zielona G\'ora, Poland }}

\pubyear{2011}
\volume{285}  
\pagerange{119--126}
\jname{New Horizons in Time Domain Astronomy}
\editors{E. Griffin,  R. Hanisch, \& R. Seaman eds.}
\begin{document}

\maketitle

\begin{abstract}
Being fast rotating objects, isolated neutron stars (INSs) are obvious targets for high-time resolution observations. With the number of optical/UV/IR INS detections now increased to 24, timing observations become more and more important in INS astrophysics. 

\end{abstract}

\firstsection 
              
\section{Introduction}
             
So far, 24 INSs 
have been  detected in the ultraviolet (UV), optical, and infrared (IR) (Mignani 2011a). A comprehensive discussion of optical timing of all INS types can be found in Mignani (2010a). Here, I focus on those classes for which counterparts at any or more of  these wavelengths have been detected.   Apart from the pulsars, these are the magnetars  (Mereghetti 2008) and the X-ray Dim INSs (XDINSs; Turolla 2009). 
Optical timing yields the direct INS identification, inputs to neutron star magnetosphere models through the comparison of the multi-wavelength light curves,  evidence of debris disks,  the spin-down parameters of radio silent  INSs, and the measurement of Giant pulses, only detected in the radio and optical bands (Mignani 2010b). 
In the following, I describe the optical pulsation emission mechanisms, the observational challenges in timing studies of INSs, and outline the characteristics of pulsations from pulsars and magnetars.
           
\section{Optical pulsations: mechanisms and observations}

The production of optical pulsations from INSs  depends on the underlying emission process. In some cases, the optical emission is the result of energy irradiation from relativistic particles in the neutron star magnetosphere through synchrotron losses or other non-thermal processes. Optical pulsations are, then, expected from INSs which have a strong magnetospheric activity, i.e. pulsars and magnetars. In this case, the optical emission is produced near the magnetic poles, yielding a small beaming factor.  Indeed, optical pulsations are mostly characterised by sharp, double-peaked, profiles. Phase shifts with respect to the  X-ray and $\gamma$-ray light curves are usually observed if the emission comes from different regions in the magnetosphere with, e.g. the $\gamma$-ray emission produced in the outer magnetosphere. 
In some other case, the optical emission is of thermal origin and produced by the cooling of  the neutron star surface. Optical pulsations are, then, expected from INSs with dominant thermal emission components, e.g. the XDINSs, if the optical emission is associated with a non-isotropic temperature distribution on the neutron star surface. In this case, optical pulsations are expected to have shallow profiles, while phase shifts between the optical and X-ray light curves are a natural consequence of the emission being produced from areas of the neutron star surface at different temperatures.
Finally, it is possible that optical pulsations do not directly originate from the neutron star magnetosphere or surface but from the reprocessing of the pulsed X-ray radiation in a circumstellar debris disk, formed out of fallback material after the supernova explosion. In this scenario, optical pulsations can be obviously produced from any type of INS with a disk. The reprocessing affects the optical light curve, with wider profiles expected with respect to the X-ray one due to the smearing of the X-ray pulse by the disk material.  Moreover, phase shifts between the optical and X-ray light curves are expected, which are due to the radiation travel time between the neutron star and the disk, and typically depend on the size of the disk inner radius and geometry. Depending on the disk viscosity, a continuum emission component can be present, which might be stronger than the pulsed one, produced by the X-ray reprocessing.  
Only 8 of the 24 INSs detected at optical wavelengths (Mignani 2011a) are also detected as optical pulsars. Of course, there are several reasons for this paucity (30\%) of detection.  The first one is related to characteristics of their optical emission and to their intrinsic faintness, which limits the search for a periodicity. Indeed, only three INSs are brighter than $V\sim25$. Moreover,  the value of the Pulsed Fraction (PF) depends on the underlying emission process and it is difficult to determine it {\it a priori} without knowing the nature of the optical emission, i.e. without adequate spectral information. This is usually obtained through multi-band photometry measurements which, sometimes, are collected over several years. At the same time, the value of the PF measured at other wavelengths, e.g. in the X-rays, cannot be taken as an absolute reference since light curve profiles vary significantly as a function of  wavelength, very much like the INS spectrum.  The slope of the optical spectrum, and/or the extinction along the line of sight, also biases the choice of the observing wavelengths which, in turns,  has to cope with the availability of a detector/instrument working in that wavelength range. Moreover, as observed, e.g. in the magnetars, the PF very much depends on the source brightness, varying significantly from active to quiescent states. 
The second reason is related to the difficulties in running periodicity search algorithms. The low number of photons which can be collected over a few hours long integrations, makes it impossible to analyse the time series through a Fast Fourier Transform (FFT).
Thus, one needs to fold the time series around a reference period available from radio or X, $\gamma$-ray observations. In this case, the INS must be a stable rotator, i.e. it must now feature sudden variations of the spin-down rate (glitches), otherwise the search for pulsations would require contemporary ephemeris. A further difficulty is in the {\it a priori} estimate of an expectation value for the PF, hence of the required integration time. 
The last reason is related to the availability and the characteristics of the instruments used for optical timing. For instance, depending on the spectrum and extinction, one can expect, for a given object, an higher flux e.g., in the IR than in the UV.  Pulsations might, thus, be undetectable outside an optimised wavelength range, which implies that there is an instrument/detector selection effect. Moreover, the timing of fainter INSs became feasible only with the advent of 8m-class telescopes, like the {\em VLT} or the {\em Gemini}, at the end of 1990s. Even in this case, however, the search for optical pulsars clashed with the paucity of on-site instruments for high-time resolution observations, being them photon counters, time-resolved imagers, or fast read-out windowed CCD devices.  The {\em HST} has been equipped with instruments for high-time resolution observations in the optical/UV, e.g. the {\em HSP} (10 $\mu$s) and the {\em STIS} (135 $\mu$s). However, the former  was removed during the first Shuttle Servicing Mission in 1993, while the latter has been unavailable between 2004 and 2009 for a technical problem, which could not be fixed due to the temporary Shuttle flight ban. On the other hand, most  timing facilities at ground-based telescopes are guest instruments, built, maintained, and operated by private consortia, and not directly available to the Community for open time proposals. Moreover, they are not easily portable, 
have to be properly interfaced to different telescopes structures, hardware, and the  instrument shipping has non-negligible costs in term of travel expenses both for equipments and man power.
So far, five of the 12 identified pulsars also pulsate in the optical. In general, pulsars are the best target
INSs for optical timing since they usually count on accurate radio ephemeris, spin-down parameters, distances, positions, with many potential targets routinely discovered in radio and $\gamma$-ray surveys. Moreover, many pulsars are observed in X-rays, which gives the interstellar extinction via the $N_{\rm H}$, hence an estimate of the brightness and of the most-suited observing wavelength. In general, optical light curves of pulsars are all double-peaked, with a phase separation $\Delta \phi =04$--0.6, with the only exception of  B0540$-$69 for which $\Delta \phi \sim 0.2$.The peaks in the optical
light curve are not always in phase with the $\gamma$/X/radio ones, as expected from the pulse origin in different regions of the magnetospheric. All pulsars but B0540$-$69 are also detected as optical/UV pulsars but only the Crab is also detected
as an IR pulsar.  Interestingly, one of the very few measurements of a pulsar braking index has been obtained from the
optical timing of B0540$-$69 (Gradari et al.\ 2011), while the Crab is the only pulsar where Giant Optical and Radio Pulses have been observed simultaneously (Collins et al., these proc.).
Three magnetars out of 6 identified in the optical/IR pulsate.
Their emission is either of magnetospheric origin, perhaps powered by the magnetic field, or produced by X-ray reprocessing in a debris disk (Mignani 2011a and refs. therein). In all cases, the profile of the optical pulsation reproduces the X-ray one. For 1E\, 1048.1$-$5937, the optical PF is $\sim$70\% of the X-ray one, providing evidence for disk reprocessing. However, there is also a marginal evidence ($2 \sigma$) for X-ray lags, not expected by the reprocessing scenario. For 4U\, 0142+61 the optical PF is larger than the X-ray one and there is evidence ($2 \sigma$) of optical lags. Finally, for SGR\, 0501+4516, the optical PF is a larger than the X-ray one and the optical light curve is in phase with the X-ray one.

\section{Future perspectives}

The wealth of pulsating INSs detected in the X ($\sim$60) and $\gamma$-rays ($\sim$80) highlight a quantitative gap between the UV/optical/IR (8) and high-energy domains. The collecting areas of the Extremely Large Telescopes (ELTs), together with new generation instruments, is needed to start a new era in optical timing (Mignani 2010b, 2011b) and match the potentials of the {\em LOFT} X-ray mission (Mignani et al., these proc). {\em QuantEye}, the first pilot study for the OWL 100m telescope, was based on quantum detector technology 
to reach pico-s time resolution (Barbieri et al., these proc). It was father to prototypes for the 
Asiago 182cm (AQuEye) and the 3.5m NTT telescopes (IQuEye), which produced the best measurements of pulsar light curves.
 A new prototype (Equeye) is being studied for the {\em VLT} (PI: Barbieri) as a possible precursor for a new quantum photometer for the {\em E-ELT}. This will open a new era in optical timing studies of isolated neutron stars.

\end{document}